\documentstyle[12pt,aaspp4] {article}

\righthead{Q0957 Structure}
\begin{document}

\title{ Accretion Disc Structure and Orientation in the Lensed and Microlensed
Q0957+561 Quasar }

\author{Rudolph E. Schild}
\affil{Harvard-Smithsonian Center for Astrophysics, 60 Garden Street, Cambridge
MA 02138}

\begin{abstract}

Because quasars are unresolved in optical imaging, their structures must
presently be inferred. Gravitational microlensing offers the possibility to
produce information about the luminous structure provided the Einstein ring
diameter of the microlensing particle is comparable to or smaller than the
radiating quasar components. Particularly interesting is the case
of multiply images gravitational lenses, where differences in the
brightness histories of the multiple images can reveal the presence of the
microlensing particles and allow inferences about the quasar's structure.

The long brightness history measured for the Q0957 quasar has been analyzed
previously for information about the microlensing particles, and evidence
for the existence of a cosmologically significant population of planetary
mass particles has been reported. The microlensing results have also
directly determined the sizes of the ultraviolet light emitting surfaces 
in the quasar

Autocorrelation analysis of the same
brightness record has produced evidence for complex structure in the
quasar; if the quasar suddenly brightens today, it is probable that 
it will brighten
again after 129, 190, 540, and 620 days. We interpret these lags as the
result of luminous structure around the quasar, and in particular we
interpret them in the context of the Elvis (2000) model 
of the quasar's structure.
We find that the autocorrelation peaks imply that beyond the luminous inner 
edge of the accretion disc, the biconic structures of the Elvis model must
lie at a radial distance of $2\cdot 10^{17}$ cm from the black hole,  
and $2\cdot 10^{16}$ cm above and below the plane of
the accretion disc. The quasar is apparently inclined 55 degrees to the
line of sight. A second possible solution with lower inclination and larger
structure is also indicated but statistically less probable.

\end{abstract}


\section{Introduction}

Structure in quasar accretion discs should be resolvable from gravitational
microlensing if the structure size is comparable to the resolution scale of the
microlensing object. We envision the typical gravitational lensing
situation such as was first seen in the Q0957+561 A,B quasar in which the
mass of a lens galaxy along the line of sight to a distant quasar creates
multiple light paths to the quasar. In the Q0957 case, two images are seen
(Walsh, Carswell, and Weymann, 1979).
Thus, measurement of brightness fluctuations
in the two images allowed the lag between the arrival times of the images to
be measured (Schild and Cholfin, 1986) and it was immediately recognized
that the brightnes fluctuations seen in the two images are not identical
(Grieger, Kayser, and Refsdal, 1988; Vanderriest et al, 1989).
This causes complications in the determination of the time delay, but has
the potential to reveal something about the nature of the quasar's
brightness distribution, and the nature of the missing mass objects in the 
halo of the lens galaxy.

Before the discovery of such microlensing brightness fluctuations,
they were predicted by Chang and Refsdal (1979), Young (1981) and Gott (1981).
Most of these calculations assumed that the luminous quasar source
consisted of a simple accretion disc with an outer diameter D and
perhaps with an unimportant central hole of unknown diameter. The most
recent calculations, by Schmidt and Wambsganss (2000), are still made with
this assumption. This is perhaps justifiable because no standard 
physical model for
a quasar accretion disc that explains all the complex observed phenomena
yet exists. 

While studies of quasar brightness have focused on the continuum emission,
typically peaking in the ultraviolet at the quasar's rest frame, it has
long been known that the broad emission lines probably originate in a more
extended region, and in order to explain the many spectroscopic phenomena
observed, various unification models have been developed. However there has
been virtually no discussion of the possibility that continuum emission
also originates in these larger regions. A further complication is that
rapid quasar brightness fluctuations are frequently seen, albeit at low
amplitude (Colley and Schild, 2003), which has tended to be interpreted 
as evidence that the accretion disc is small.

What seems to be lacking in the discussion is the possibility that the
quasar continuum emission comes from several regions, large and small, with
smaller regions producing rapid brightness fluctuations as observed and
larger regions producing the pattern of multiple peaks in the brightness
curves observed. The multiple peaks are evidently produced when the central
black hole's variability produces brightness enhancements at the several
continuum emitting regions, which are seen at different times (lags) by the
terrestrial observer. A double-ring quasar model by Schild \& Vakulik
(2003) has reproduced all the complex microlensing phenomena observed to
date, and will be the basis for the further developments in this report.
Note the potential for confusion in referring to the ``double ring model.''
The Elvis et al (2000) model has conic sections of revolution which produce a
luminous ring above and below the accretion disc, whereas the Schild and
Vakulik model has a small ring to represent the luminous inner edge of the
accretion disc, and an outer ring which is a simplification of the two
rings in the Elvis model. Hereinafter we refer to the Schild and Vakulik
model as the double ring model and the conic section model as the Elvis
(2000) model.

Extensive brightness monitoring of the Q0957 quasar for time delay and
microlensing studies has produced four relevant observations of quasar
structure so far. First, the microlensing record that results from
subtracting the time delay shifted brightness record of the first arriving
A image from B produces a record of an event with an amplitude of 30 \% and
lasting 10 years, followed by a 10-year period with no secular increase on
such time scales (Schild \& Smith, 1991; Pelt et al 1998, Refsdal et al
2000). Microlensing simulations (Young et al 1981; Kayser, Refsdal, and 
Stabell, 1986)
have always suggested that the timescale for a solar mass star crossing the
luminous quasar structure should be approximately 30 years, based upon the
expected diameter of the microlens' Einstein ring. Of course stars
are presumed to exist in the outer portions of lens galaxy G1 since their
light and spectra are observed, so the long secular microlensing drift is
presumed to result from microlensing by stars.

A second and unexpected observational result is the more rapid microlensing
reported in Schild (1996); cusp-shaped profiles having an amplitude of 5 \%
and a time scales of 90 days were found and interpreted as indicative of a
microlensing population of dark matter objects of planetary mass,
$ 10^{-5} M_{\odot}$,
and called ``rogue planets, likely to be the missing mass.'' The Schild \&
Vakulik model demonstrates how these profiles can originate in the strong
shear of the solar mass microlenses provided the quasar has sufficiently
sharp structure. The even more rapid event, albeit at lower $1\%$ amplitude,
reported by Colley and Schild (2003) was of the type reported by 
Schild (1996) to
be indicative of even lower mass microlenses, $10^{-6} M_\odot$, although
very sharp quasar structure would then be implied.

A third important conclusion from the microlensing record is that no
fluctuation of more than a factor of two has ever been recorded in Q0957
or any other gravitationally lensed quasar, although microlensing of stars
in the LMC and galactic center have frequently produced such spectacular
brightness increases (Alcock et al 1995).
An analysis by Schild (1996) of the Q0957 brightness record shows that the
quasar images have only produced fluctuations of 0.45 magnitudes 
in 100 years of brightness monitoring. Analysis of the single
event occurring on solar mass microlensing timescales by Refsdal et al
(2000) shows that large quasar sizes are implied. A general
theory of quasar brightness fluctuations has been published by Refsdal and
Stabell (1991, 1993, 1997) and applied in Schild (1996) to the Q0957 system
to show that the quasar luminous source must be at least as large as 
$10^{16}$cm, and probably larger.

A fourth important inference, which is the starting point of this paper,
comes from the quasar brightness curves
directly, with microlensing playing only a minor role. Autocorrelation
calculations of the brightness records show a network of peaks on time lags
of order 200 proper days, suggesting that the quasar's luminous structures
are at
size scales of 200 light days, and that the source of this luminosuty is
probably scattering of the emitted continuum. 
Of course this seems to be contradicted by
the observations of brightness fluctuations on 1 day time scales as
mentioned above, but because the fluctuations have low amplitude it is
possible that the quasar luminosity comes from several regions with size
scales between 3 days and 200 days (Schild and Vakulik 2003). It is then 
necessary to ask what the
effect of microlensing would be on such a large luminous object. We recall
the important estimate that the optical depth to microlensing for the
lens system is (0.34, 1.25) for the (A, B) image (Refsdal et al, 2000). 
This means that for any
point on the luminous quasar surface, the probability is greater than 1
that it is microlensed, and that on average 1.3 objects contribute to
the microlensing of image B. 
Because the A and B images presumably lie behind
different patterns of microlensing cusps originating in the lens galaxy, it
would be expected that the quasar's luminosity distribution would be
differently microlensed in the two images, and that since the different
quasar luminous regions arrive at different times over a time interval
around 200 days and are selectively amplified, the two quasar images might
differ in their brightness autocorrelation properties, as observed.

The purpose of this contribution is to consider what is known about
possible structure in the autocorrelation properties of the brightness
curves of the two quasar images, and to show how the structures indicated
can be related to the predictions of the best available quasar model.
The Elvis (2000) model finds that in addition to the accretion
disc, a pair of conic section surfaces of revolution
must be present to explain the
observed emission and absorption line features long recognized in quasar 
spectra.  These ``biconic surfaces'' are Compton thick 
where illuminated by the black hole,
and so they would be expected to scatter continuum radiation into our line
of sight. We find
that such structure predicts the pattern of $\sim 200$ day lags detected in
the extensive Q0957 monitoring. Seen as a system of 4 equations and 3
variables, the problem is overconstrained and allows us to determine 
the inclination and dimensions  of the quasar.

In section 2 we show that the first arriving
optical signal probably arrives from the inner edge of the accretion disc,
which evidently has a proper size scale of approximately 3 light days, 
as estimated
from the autocorrelation data. In section 3 we identify the autocorrelation
peaks suggesting the existence of larger quasar structure. In section 4 we
show that the autocorrelation peaks can be interpreted as originating in
the biconic luminous structures found in the Elvis (2000) quasar model, and we
determine the sizes of these structures and their orientations. We
summarize our conclusions in section 5.

\section{ The Inner Edge of the Accretion Disc}

Extensive records of the brightness of the two Q0957 images have been
compiled by Schild and colleagues; see Schild and Thomson (1995) and earlier
references contained therein. A master data set is available at:
           http://cfa-www.harvard.edu/~rschild/fulldata2.txt
This data set was analyzed by Thomson and Schild (1997) and Schild and
Thomson (1997 a,b)
with the striking discovery that significant power is seen at four lags
which we identify as T1...T4 in Fig. 1. An additional unexpected behavior
was also seen for very short lags, of order 10 days (observer's clock), or
4 proper days for a quasar redshift of 1.41. Although not really
well seen on Fig. 1, the autocorrelation has a small secondary peak at 10
days in the A image, and a different, more complicated structure, in the B
image.  Note that the Schwarzschild radius
for a $3\cdot 10^{9} M_\odot$ black hole would be $4.4\cdot 10^{15}$ cm, 
so the diameter of the
innermost stable orbit is $10^{16}$cm, or about 2 light days, 
which is the size
of the smallest structure inferred from the autocorrelation. We suspect
that the differences between the two images results from microlensing of
the luminous inner accretion disc edge.
Because the
quasar's brightness has been sampled nightly, and assuming a finite
thickness to the emitting region at the inner edge of the accretion disc,
it would be unsurprising if the microlensed inner edge of the accretion
disc contributed some signature to the autocorrelation estimates on time
scales of a few days. Therefore we plan to investigate more completely the
behavior on short time scales in future work by sub-sampling of the long
Q0957 brightness record into 1-year subsets.
For the present report we
take the size of the luminous accretion disc inner edge to be a few proper
light days. Following Schild \& Vakulik (2003) and Colley et al (2003) we
estimate that approximately 20 \% of the quasar's continuum originates there.

\section{ The Measured Time Lags}

Our fig. 1 autocorrelation plot for the quasar brightness history was
calculated by the multiple window technique as described in Thomson and
Schild (1997). Further discussion of the statistical complexiteis of the
Q0957 data are given in Schild \& Thomson (1997 a,b) particularly with 
reference to non-Gaussianity in the basic statistica processes. 
An important shortcoming to Fig. 1 is the absence of error bars or
confidence limits on the plotted curves. To date this has been a result of
lack of understanding of the basic statistics of the
quasar's brightness fluctuations.
Although confidence limits can be estimated for Gaussian statistics,
it is well known that the basic statistics are not Gaussian, as noted above 
(Thomson and Schild, 1997; Press and Rybicki, 1997). 
Although the latter sought the
non-Gaussianity via asymmetry of the brightness peaks, Thomson and Schild
have concluded that the non-Gaussianity lies in the cuspy profiles
resulting from microlensing. Relative to a Gaussian distribution, the cuspy
profiles produce too many 2 sigma to 5 sigma discrepant points, as may be 
seen from a plot of the higher statistical moments of the brightness record.

One might imagine determining limits to the errors in estimating the
autocorrelation by comparing the autocorrelation function for the two quasar
images; after all, they are images of the same quasar. However this would
fail, for the reason that the different quasar images are seen through 
different
microlensing cusp patterns originating in the lens galaxy, and the cusp
patterns will magnify different parts of the quasar's brightness
differently. Note that for lags greater than 600 days, the estimates for
the two images differ by a factor of two over a wide range of lags, and
such a result is unlikely to result from errors of estimation. As expected,
it is the B
image with the greater optical depth to microlensing that shows the
stronger autocorrelation estimates for longer lags.

Until the statistics of the basic brightness curves are better understood
we simply take the autocorrelation estimates at face value and attempt to
interpret their peaks. We have identified the most prominent four peaks
that are coincident in both the A and B image brightness records, and we
attempt to interpret them in the context of the Elvis (2000)
model of the luminous quasar structure. In this model, the quasar is
expected to have an accretion disc, which does not need to be thick, and
two biconic figures of revolution, one above and one below the accretion
disc. The many optical, UV, and X-ray spectral lines seen in emission 
and absorption are
understood to originate in these biconic structures. Quasars of different
classes, like the BAL quasars and weak absorption line quasars, are
understood to result from the different orientations in which the
structures are viewed from earth.

Of particular importance to us, since we are analyzing brightness records
of the quasar's continuum emission, are the regions scattering the central
continuum. According to the Elvis model, these will be two approximately
circular rings at the $\tau \sim 1$ surface, one above and one below the 
accretion disc. The reader is referred to Elvis (2000) Fig. 1
for a cartoon illustrating these structures. 
Since in general the quasar axis will be inclined 
to our line of sight, light scattered from these surfaces will be received at
Earth at different times. Thus different parts of the rings of Saturn would
be seen at different times if one imagined a strobe light at the position
of the center of the planet flashing very quickly. And as in the Saturn
example, the light scattered from ring portions seen in projection nearest 
to the strobe
light would have the shortest and longest lags. These topics have been
covered extensively in Peterson (1997). The segments
with the longest and shortest lags would be the brightest. In a similar
way, we expect the surfaces of the Elvis model biconic structures to be
most luminous in the segments having the longest and shortest lags.

We have identified the four principal lags T1...T4 in figure 1 from simple
visual inspection. We sought the four regions displaying the most prominent
peaks in both the A and B image. We were biased away from the feature at
365 days which could too easily be an artifact of sampling. We ignored the
many prominent peaks seen in one image but not the other, like the peaks at
observed lags of 60 days and 240 days in the A image. These could easily be the
result of some locally strong microlensing of some portions of the emitting
circular structures, where the lags are 60 and 240 days (25 and 100 proper
days). The strongest lags in the autocorrelation plot are at 50, 75, 230,
and 260 proper days (in the source plane time frame). The 260 day peak is
marginal. The peaks are resolved with widths of 15 proper days.

We thus seek to identify these four most prominent peaks, labeled T1 ... T4
in Fig. 1, with the near- and far-side emitting regions of the biconic
sections (rings). Two simple configurations are possible for 
identifying the regions,
which we call case 1 and case 2. Case 2 is the less probable,
given that it requires that the pole of the rotating accretion disc be almost
coincident with the line of sight to Earth. The pole of rotation is more
inclined in case 1, and so case 1 is statistically 
more probably given that the
line of sight to earth and the pole of rotation are presumably random
vectors.

Figure 2 is a cartoon showing the luminous quasar structures; we
define the variables chosen to describe them for both case 1 and case 2.
The biconic surfaces and the
accretion disc are shown in their principal cross sections. In both cases 
we suppose that a disturbance near the black hole event
horizon causes energetic emissions that are seen first as brightenings of
the inner edge of the accretion disc, through some process of scattering or
fluorescence. We then imagine that this starts our clock at t = 0, whence
the radiation spreads to the biconic surfaces and causes them to brighten
at appropriate lags; we also assume that the conversion to the radiation we
detect (electron scattering in the Elvis 2000 model)
occurs on time scales short in comparison to the few days of the
geometric lags. Then from the simple geometry shown in Figure 2, we easily
write the equations for the predicted lags, as functions of the geometric
variables r, the distance in units of proper light days
from the central black hole to the luminous
bi-conic structure, $\theta$, the angle between the line of sight and the
pole of the accretion disc, and $\epsilon$, the quasar structure variable
describing the angular height of the luminous rings above the accretion
disc as seen from the black hole:

	$t1 = r(1-sin[\theta + \epsilon]) = 50$

	$t2 = r(1-sin[\theta - \epsilon]) = 75$

	$t3 = r(1+sin[\theta - \epsilon]) = 230$

	$t4 = r(1+sin[\theta + \epsilon]) = 260$

Here we have 4 equations for 3 variables, so the system is over-constrained,
and it is not assured that a solution can be found. On the other hand we
lack good limits on the precision of the measured lags, so we do not
concern ourselves about the existence of a solution. Particularly
problematic is T4, where the lag peak differs somewhat for images A 
(265 days) and B (257 days).
In our calculation, we have used the average lag for T4.

Our case 1 differs from case 2 in the order in which the lagged signals
from the different emitting structures arrive at earth. In both case 1 and
2, the surface nearest the earth is seen first. In case 1 the luminous
surface behind the accretion disc from the first emitting surface is seen
second; in case 2 it is seen third.

Our solution for the statistically more probable case 1 is given by:

	$[ r , \theta, \epsilon ] = [ 154, 36, 6.5 ].$

Here r is in units of light days, and the angles are expressed in degrees.

For the less probable case 2 the equations are again easily written with
the same variables:

	$t1 = r(1-sin[\epsilon + \theta]) = 50$

	$t2 = r(1-sin[\epsilon - \theta]) = 75$

	$t3 = r(1+sin[\epsilon - \theta]) = 230$

	$t4 = r(1+sin[\epsilon + \theta]) = 260$

Our case 2 solution, expressed in the same units as for case 1 is:

	$[ r, \theta, \epsilon ] = [ 154, 7, 36 ].$
	
Thus far our discussion has treated the autocorrelation peaks as being delta
functions and treated the circular scattering regions detected as
very thin and well defined. In fact, the autocorrelation peaks have a mean
Full Width at Half Maximum of 15 proper light days. Thus we compute that
the radial extent of the scattering rings is of order 15 light days, or 
$4\cdot 10^{16}$ cm. Such estimates neglect inclination and other effects, and are
likely to be uncertain by a factor of two.

The cartoon in Figure 2 has been drawn to approximate scale for the
case 1 solution determined above.

\section { Comparison with Other Estimates of Quasar Size}

Some inferences about size of the continuum emitting region in the Q0957
quasar have previously been advanced by Schild (1996). These estimates
result from application of the Refsdal and Stabell (1991, 1993, 1997) theory of
the amplitudes of the observed brightness fluctuations, applicable in the
case of large quasar accretion discs. Schild showed that the Q0957
accretion disc is large enough for the Refsdal-Stabell
theory to be
applicable, and gave an estimate of the size
of the luminous region, expressed as the diameter of a luminous disc if it
is assumed that the luminous source is such a structure. In fact, the theory
applies to the present geometry just as well as for a solid disc of uniform
luminosity as envisaged by Refsdal and Stabell, as long as the dimensions
of the structures found are comparable to or larger than the Einstein ring
of a microlensing particle. This condition is satisfied here; the radius
and thickness of the luminous rings of the biconic structures in the Elvis
model are $2\cdot  10^{17}$ cm and $2\cdot 10^{16}$ cm, whereas the 
Einstein Ring radius
for a half solar mass microlensing particle quoted in Schild (1996) is
$2\cdot 10^{16}$ cm. Thus it would be expected that the area of the emitting
region in the Refsdal-Stabell theory and from our present geometrical
size scales should be in approximate agreement.

We start with an estimate of the area of the luminous outer ring as estimated
above from geometric factors inferred from autocorrelation peaks and
widths. Computing area $A = 2\pi r\Delta(r)$ with r = 154 light days or $2\cdot
10^{17}$ cm and $\delta(r) = 2\cdot 10^{16}$ cm, 
then the total emitting area is $2.4\cdot10^{34} cm^{2}$. 
Since two rings are present, the total luminous area 
estimated from geometric factors is $5\cdot10^{34} cm^{2}$.

The above estimate can readily be compared with the microlensing based
estimate that follows from application of the Refsdal-Stabell (1991, 1993,
1997) theory as previously discussed by Schild (1996); from Schild's
diameter estimate we readily compute an estimate of the luminous area as 
$6\cdot 10^{34} cm^{2}$. We consider the
two results to be in reasonable agreement, given that the emission from the
luminous inner edge of the accretion disc contributes to brightness
fluctuations but has been ignored, and we have assumed that the luminous
rings in the Elvis model radiate isotropically. 
Thus we find two consistent area 
estimates of the quasar's luminous structure to be approximately 
$5\cdot10^{34} cm^{2}$.

We conclude that the data from quasar brightness monitoring consistently
indicate large sizes in the continuum emitting structures. This was
already anticipated by Schild(1996) when he first computed the quasar size
from the brightness fluctuations and noticed that the structure inferred
from autocorrelation peaks was larger by a factor of 10; he concluded that
``the quasar's luminous source is structured, and consists of rings or
clouds with a $10\%$ filling factor.'' The further development of this
conclusion has produced the Schild \& Vakulik (2003) model that can
reproduce all the microlensing phenomena observed to date.

\section { Summary and Conclusions}

We have considered the implications of the autocorrelation analysis 
of continuum brightness
fluctuations in the Q0957 quasar and found that structure on time scales of
200 light days is indicated. We have examined how peaks in the
autocorrelation function for the two quasar images can be interpreted in
the context of the Elvis (2000) model of quasar structure to determine the
sizes and positions of the luminous structures, and perhaps the orientation
of the quasar to our line of sight. Because the rotation axis is presumably
aligned with the 31 degree east-of-north radio jet (Roberts et al 1985)
we can fully specify the orientation of the quasar in 3 dimensions.

We do not consider that our analysis of the brightness record proves that
the Elvis model is correct. On the one hand, the identification of four
peaks produces an overconstrained problem which nevertheless produces a
solution that agrees with other estimates of the quasar's size and in
particular the unique parameter $\epsilon$ that describes the height of the
rings above the equatorial plane (accretion disc). On the other hand one of
the four autocorrelation peaks is weak (probably for a good reason), and
it might be argued that the four correlation peaks have been rather
arbitrarily identified with the quasar's structure for a particular model.
Nevertheless we believe that there is compelling evidence for
the existence of some kind of structure on proper time scales of 100 light
days, that affects a significant fraction of the quasar's luminosity.
Other structure on time scales of only several light days is probably also
present, and we presume this to be the luminous inner edge of the accretion
disc. 

The existence of 2 distinct luminous regions in quasars has long 
been inferred from
quasar spectral energy distributions (Elvis et al 1994 and earlier
references contained therein).
Such energy distributions have long
been characterized by a power law over a large spectral band extending from
approximately the Lyman limit at 1216 angstroms to the near infrared. 
The biconic surfaces are probably responsible for the power
law continuum and cover a radial distance range of 
approximately $2\cdot 10^{16}$
cm. This is also probably the region where the emission lines are emitted,
as inferred above. In
addition a ``blue bump'' has also been recognized as a feature in the
ultraviolet, peaking at approximately 3000 Angstroms. We attribute the 
blue bump to the approximately thermal radiation emitted by the luminous 
inner edge of the accretion disc, which is probably nearly homogeneous in
temperature. The blue bump emitting region had previously been
recognized as a spatially distinct region, because its brightness has been
found to fluctuate independently of the power law continuum, and its
probable partial origin in Fe II emission lines has been noted (Wills,
Netzer, and Wills, 1985).

Our determination of luminous quasar structure on size scales of 154 light
days is probably consistent with the the Peterson et al (1991, 1993)
determination of
size in the emission line forming region of Seyfert galaxy NGC 5548.
From their monitoring of brightness fluctuations in the continuum and in
the emission lines, these authors concluded that emission lines originate
in a region 15 proper light days from a luminous continuum source,
presumably the luminous inner edge of the accretion disc. The smaller
inferred sizes of structures in NGC 5548 are probably consistent with a lower
mass central black hole. The Elvis model predicts that emission lines
originate on the bi-conic surfaces of revolution above and below the
accretion disc for NGC5548 and for Q0957, and thus it is likely that the
same volume of gas producing emission lines in the quasar is also producing
the non-thermal continuum, and a characteristic size scale and distance 
from the center has apparently now been measured.

The existence of multiple emitting regions probably also explains how the
quasar can exhibit ultraviolet brightness fluctuations on very short time 
scales of approximately a day. These rapid brightness fluctuations have
long been recognized in quasar optical brightness records (Sandage 1964)
and were first reported in Q0957 by Vanderriest et al (1982) and were
already evident in the contemporaneous data of Schild and Weekes (1984). 
They could most
easily arise in the smallest quasar structure we identify, the inner edge
of the accretion disc. We have already shown that the innermost stable
orbit in the accretion disc is only 2 light days from the center of the
black hole for our estimated black hole mass, and our observations of the
autocorrelation lags for the inner structure suggest a diameter of around
10/2.41 light days, in agreement with expectations. 
As emphasized by Schild (1996, 1999) the failure of time
series analysis to easily find a cross-correlation peak with a time scale
of 1 day, the normal sampling frequency, suggests that microlensing of
these small inner structures must be important. For the normally estimated
transverse velocity of the microlensing screen originating in the lens galaxy,
1000 km/sec, microlensing particles of $10^{-6} M_{\odot}$ 
would cross the luminous
structure in a day. Recall that the continuously observed microlensing is
measured at an extremely low amplitude, only .01 magnitudes, on daily time
scales (Schild, 1999, Colley \& Schild, 2003). 
Note that because the quasar structure might be
resolved by the Einstein Ring of such small microlenses, the brightness
fluctuations might not have a strongly cusp-like character. This is largely
what is observed; Schild (1996, Fig. 2 and Fig. 5) has demonstrated a 
sharply cusp-like
microlensing pattern on 90-day time scales, but the more rapid
fluctuations, with day time scales, seem less peaked (Colley and Schild, 2003).

It is also useful to recall that the radio emission has a component
originating in some region with dimensions comparable to those found
here. Cross-correlation of the radio brightness records measured at 6 cm
with the optical brightness records has shown a time lag of approximately 30
days measured, or 12 proper days (Thomson and Schild, 1997; Schild and
Thomson, 1997a). The radio emission presumably originates
in a region above (and below) the accretion disc, and evidently is a
response to the same disturbances responsible for the ultraviolet
brightness fluctuations observed at 6500 Angstroms. Because the radio
emission presumably originates in ionized gas being accelerated away from
the black hole, it is not surprising that the ultraviolet radiation from
the inner edge of the accretion disc is seen first, and our autocorrelation
lag of 12 proper days probably represents the difference between 
light travel time
to the accretion disc and to the polar region of brightest radio emission.
Thus the radio bright region centers 16 proper light days above the balck
hole and accretion disc.

It does not escape our attention that the Elvis structures above and below
the accretion disc plane are not a common feature of standard quasar
models. However, the mass scale invariant rotating magnetic models of
Romanova et al (2002, Figs. 16, 17; 2003, Figs. 2, 3) feature such outflow
structures. These Romanova et al simulations also feature an inner
ring-like structure similar to the inner ring of the Schild \& Vakulik
(2003) empirical model. Whereas the Romanova et al models are
non-relativistic, the Elvis structures they feature are produced in the
non-relativistic outer region and are presumably relevant in galactic black
hole candidate objects and in quasars. General relativistic effects will
predominate in the region of the inner ring of the Schild \& Vakulik
(2003) model, and the fully relativistic ``magnetic propellor'' models of
Robertson \& Leiter (2002, 2003) feature such an inner ring structure.
Such models, when scaled up from stellar to quasar masses, imply that the
quasar MECO operates in a low-hard spectral state.

\newpage

\begin{figure} [t]
\begin{center}
\plotone{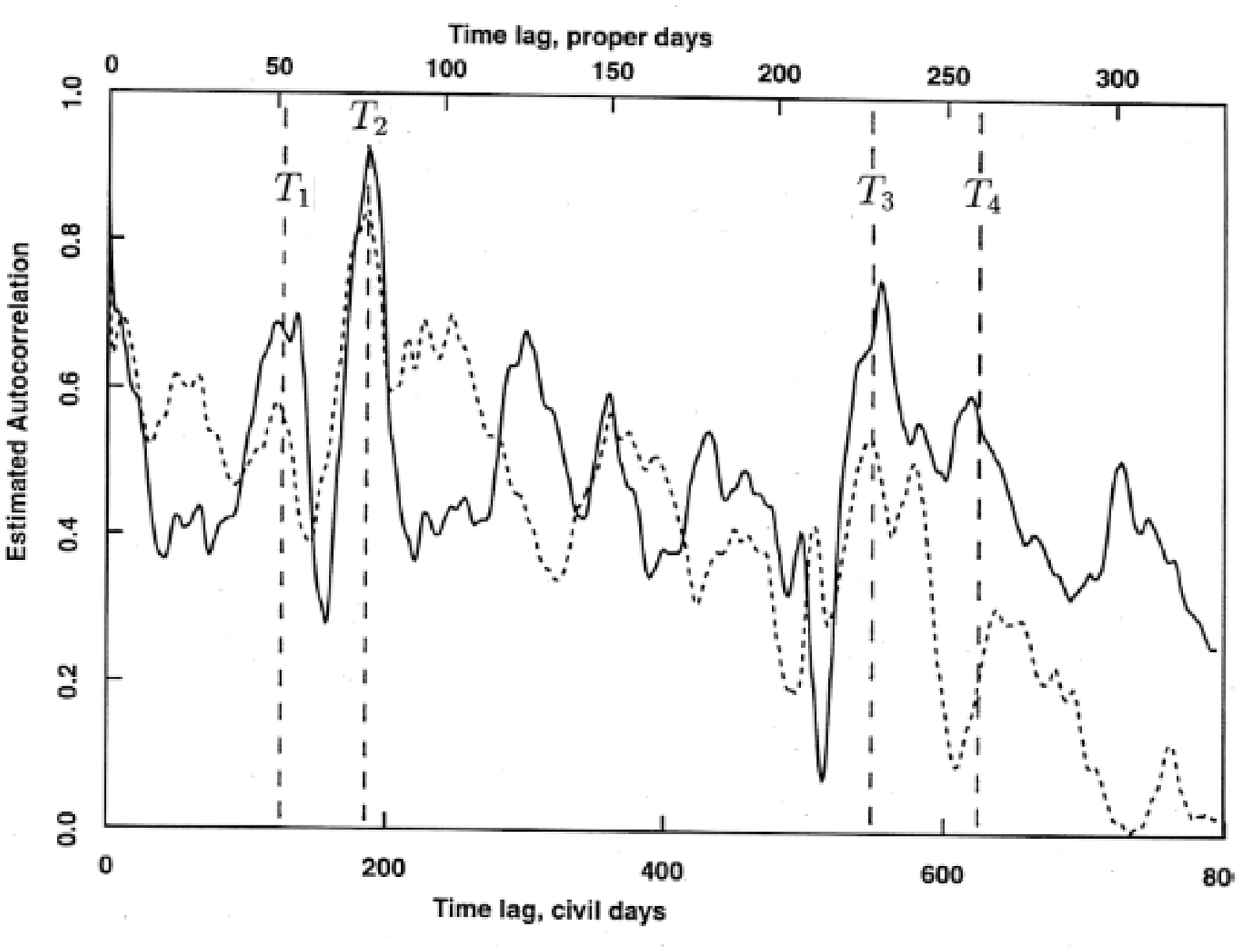}
\caption{ The estimated autocorrelation for the A (dotted curve) and B
(solid curve) quasar images, from Thomson and Schild, 1997. We have
identified the four coincident autocorrelation peaks as $T_{1} ... T_{4}$.
Not well illustrated is the structure for lags of up to 10 civil
days. Although autocorrelation estimates should be accurate for such lags
because of the large amounts of data with appropriate time sampling, the
autocorrelation estimates for the A and B images are in disagreement, apart
from the four lags $T_{1} ... T_{4}$ .}
\label{fig1}
\end{center}
\end{figure}

\newpage

\begin{figure} [t]
\begin{center}
\plotone{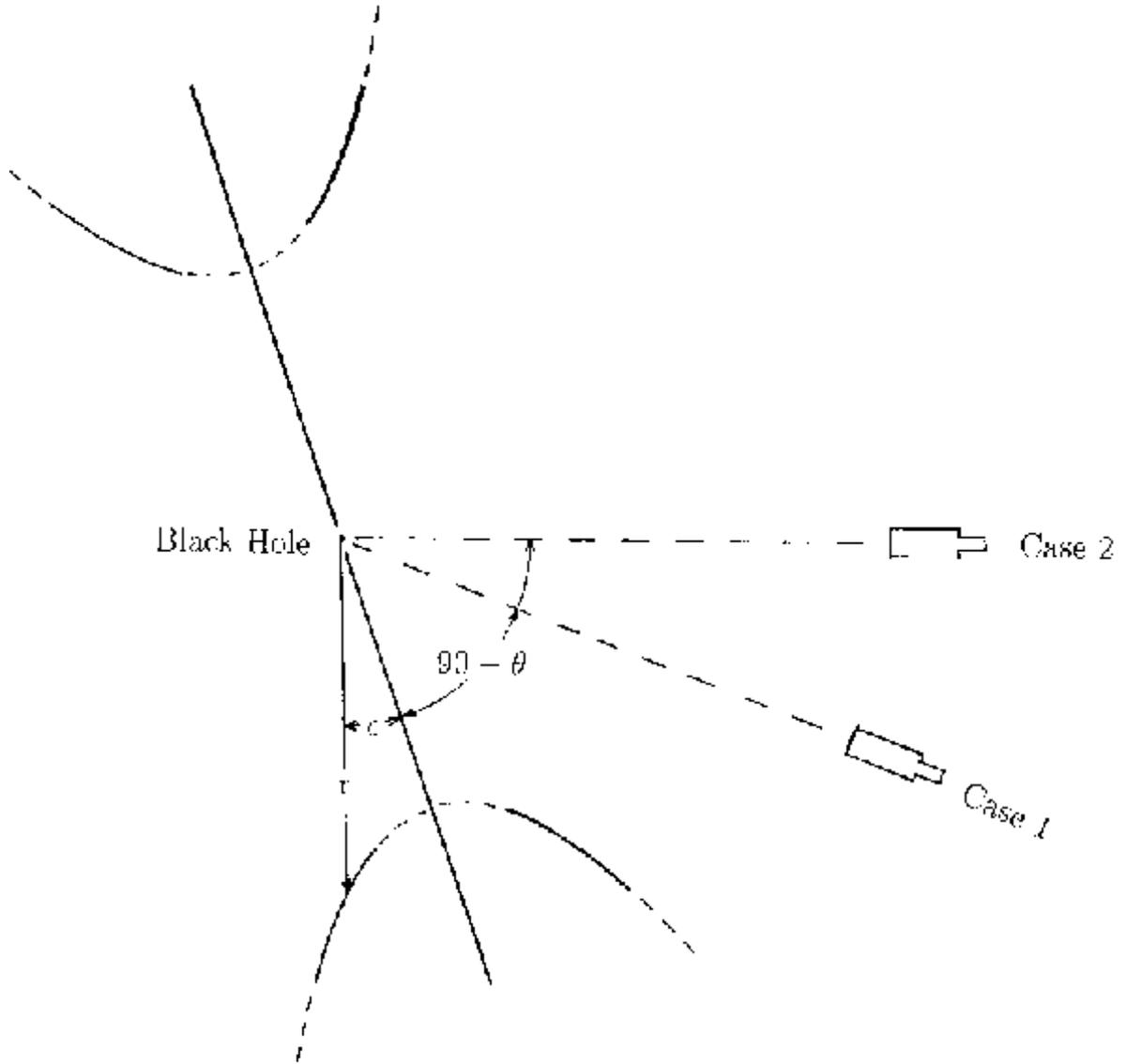}
\end{center}
\caption{ A cartoon showing the principal luminous quasar structures, and
identifying the variables used in fitting to the autocorrelation features. 
The structures are shown approximately to scale, but the inner edge of the
accretion disc must have a higher surface brightness. The viewing angle for
Case 1 and Case 2 are also shown}
\label{fig2}
\end{figure}

\end{document}